\documentclass[12pt,preprint]{aastex}
\usepackage{graphicx}

\shorttitle{BAT AGN IR/X-ray Properties}
\shortauthors{Mushotzky et al.}

\begin{document}
\title{ Correlations of the IR Luminosity and Eddington Ratio with a  Hard X-ray Selected Sample of AGN }

\author{Richard F. Mushotzy\altaffilmark{1}, Lisa M. Winter\altaffilmark{2}, Daniel H. McIntosh\altaffilmark{3}, and Jack Tueller\altaffilmark{1}}
\email{richard@milkyway.gsfc.nasa.gov}

\altaffiltext{1}{ NASA Goddard Space Flight Center, Greenbelt, MD}
\altaffiltext{2}{University of Maryland, College Park, MD}
\altaffiltext{3}{University of Massachusetts, Amherst, MA}
\begin{abstract}
We use the SWIFT Burst Alert Telescope (BAT) sample of hard x-ray selected active galactic nuclei (AGN) with a median redshift of 0.03 and the 2MASS J and K band photometry to examine the correlation of hard x-ray emission to Eddington ratio as well as the relationship of the J and K band nuclear luminosity to the hard x-ray luminosity. The BAT sample is almost unbiased by the effects of obscuration and thus offers the first large unbiased sample for the examination of correlations between different wavelength bands. We find that the near-IR nuclear J and K band luminosity is related to the BAT (14 -- 195\,keV) luminosity over a factor of $10^3$ in luminosity ($L_{IR} \approx L_{BAT}^{1.25}$)and thus is unlikely to be due to dust.  We also find that the Eddington ratio is proportional to the x-ray luminosity. This new result should be a strong constraint on models of the formation of the broad band continuum.

\end{abstract}

\section{Introduction}
The estimate of black hole masses and the ratio of luminosity to the luminosity derived from
the black hole mass (the Eddington ratio) are the fundamental parameters in the evolution of active galaxies.   The Eddington ratio is defined as L/L$_{Edd}$, where
\begin{math} L_{Edd}=(M_{BH}/M_{\odot}) \times1.3 \times 10^{38} erg\,s^{-1}\end{math} or the limit where gravitational and radiation pressure are balanced.  It is important because it is a measurement of the ratio of energy output of a source to the
physical limit of energy output.  Understanding the energy output of active galactic nuclei (AGN) is essential to understanding both
the nature of AGN sources and the effect they have on their environment.
Therefore, measuring both M$_{BH}$ and the Eddington ratio are the subject of much recent work
\citep{kol06, net07, wu02}, based primarily on optical samples and the combination of x-ray and IR samples. These papers have either used data from the literature \citep{wu02}, the SDSS \citep{net07}, or deeper optical, x-ray and IR data \citep{kol06} to obtain conclusions about the relationship of the luminosity to black hole mass and its evolution across cosmic time. For example, \citet{kol06}  conclude that  there is strong evidence that supermassive BHs gain most of their mass while radiating close to the Eddington limit,  and that the fueling rates in luminous AGNs are determined by BH self-regulation of the accretion flow rather than galactic-scale dynamical disturbances.  

Whereas previous samples are strongly biased against the most numerous AGN in the universe, those with high line of sight column densities, e.g. `obscured' AGN \citep{mat06}, our BAT sample, since it is selected solely based on 14--195 keV flux, is not. However the BAT sample, as opposed to most optical samples, has a very low median redshift (0.03) and very few objects at $z > 0.1$.  Since there is a strong relationship between  line of sight obscuration and intrinsic luminosity \citep{toz06, mar05}, such that the less  luminous sources are much more likely to be obscured, the optical samples, which miss most of the obscured sources, are strongly deficient in low luminosity sources.  Furthermore, there is often a complex relationship between measures of obscuration in the x-ray,  IR, and optical bands and there may be strong evolutionary effects such that the fraction of obscured objects is higher at higher redshifts \citep{laf05}.  If there is a physical relationship between luminosity or stage in the life of the AGN  and the occurrence of obscuration, as predicted by several models \citep{hop06}, then the results from such surveys may be rather different than The BAT sample which is heavily weighted to the local universe with a mean redshift of $\approx 0.03$.

   In addition, there has been no conclusive determination of  the origin of the `non-thermal' continuum in AGN. In the x-ray band, there are two mechanisms that have received extensive discussion: thermal Comptonization of photons produced by the accretion disk from a population of high energy particles, perhaps in a corona above the disk \citep{zdz9x}, and emission due to jet processes
\citep{kor06}. However, there is at present no agreement as to the origin of `non-thermal' IR and optical emission in radio quiet objects or even whether it exists. Most models for the origin of the optical emission have focused on disk models (e.g. \citet{sun89, cze87}) with some Comptonization.  Dusty torus models \citep{pie93}, as the origin of the K band flux, have received support from reverberation observations \citep{sug05} in the optical (UBV) and near-infrared
(JHK) bands.  These studies found a clear time-delayed behavior of the K band with respect to the V band, strongly suggesting that the bulk of the K-band emission
originates from thermal emission of dust grains in an optically
thick torus \citep{gla04}.
There are strong correlations between the optical and x-ray properties \citep{bor0x, bro06} and the IR properties \citep{alo0x}, whose origin is not easy to explain if these continuum components originate from completely different physical mechanisms as is predicted by the most modern dusty torus models of the IR continuum \citep{nen08}.

In this paper we present the first results on the Eddington ratio from an unbiased survey of AGN selected in the 14 -- 195\,keV band with the Swift BAT survey. We find a strong correlation of the Eddington ratio to the hard x-ray luminosity over a wide range in L/L$_{Edd}$ ($10^{-4}$ to $3$) in our sample, as opposed to the narrow range in the sample of \citet{kol06} selected from the AGES survey. In addition we find that the luminosity of the point-like component in the 2MASS J and K band data is almost linearly correlated with the high energy luminosity indicating that these two components are very closely coupled  and probably have the same physical origin. Since the x-ray emission is not due to dust processes, these results indicate most of the near IR is not either.

\section{Survey}
The BAT 9 month survey \citep{tue07} is an all sky survey in the 14 -- 195\,keV band \citep{mar05} with a sensitivity threshold that varies from $1-3\times10^{-11}$\,erg\,cm$^{-2}$\,s$^{-1}$ in the 14 -- 195\,keV band, as a function of sky exposure. At $|b|> 15$\degr, the survey is almost complete with 130 total sources of which 103 are  identified AGN  and the rest are galactic objects with only 1 unidentified source. Of the 103 AGN, 14 are Blazars and will not be included in the following discussion. All but 3 of the identified objects have redshifts. The survey is continuing and the sensitivity is increasing roughly as the  square root of time \citep{tue07}.

\section{Black Hole Mass Estimation}
While there are several published papers on scaling black hole masses to optical and IR luminosities, we have chosen to use the recent paper by \citet{nov06}. They give scaling relations between the central black hole mass and global galactic properties based on optical and IR data. Since, for our sample, the only uniform optical or IR data set is the 2MASS data and we wish to minimize the contribution of possible AGN light to the total IR light, we use J and K band magnitudes. As is well known, the ratio of the host galaxy to the QSO photon flux is at a maximum
in the NIR, because the SED of many stellar populations has a maximum at 1.6\,$\mu$m and the AGN
flux has a minimum at 1.2\,$\mu$m \citep{elv94, ric06}. The AGN flux increases at shorter wavelengths due to accretion-disk or power-law emission and at longer wavelengths due to thermal emission of hot dust heated by the AGN and/or starformation contributions..

We obtained the total J and K 2MASS magnitudes (m$_J (tot)$ and m$_K (tot)$) for the BAT AGN sample, as listed in NED.
Additionally, we used the IRAF task {\tt qphot} to obtain circular aperture photometry of the
central region, equal to the PSF FWHM of the 2MASS images.  While we label these values as
the nuclear magnitudes (m$_J (nuc)$ and m$_K (nuc)$), they are clearly overestimates to the nuclear flux, since no provision is made for the contribution of the galaxy light in the central region.
We have found that in most objects the possible contribution to the total J band light from the nucleus is small and in most cases is well measured. This is important, since if the J band nuclear light dominated  the emission, our estimates of BH mass would be severely impacted and the scatter in relations involving the BH mass (such as the Eddington ratio) would have much more noise.

From the apparent magnitudes, we computed absolute magnitudes (M$_J (tot)$, M$_K (tot)$, M$_J (nuc)$, and M$_K (nuc)$)
with the redshifts listed in \citet{tue07} and assuming H$_0 = 75$\,km\,s$^{-1}$\,Mpc$^{-1}$.
From the total and nuclear absolute magnitudes, we computed estimates of the
galactic/stellar component of the NIR emission (M$_J (stellar)$ and M$_K (stellar)$) as
\begin{math}M_{stellar} = 2.5\log(\chi/ (\chi-1)) + M_{tot} \end{math} where
\begin{math} \chi = F_{tot}/F_{nuc} = 10^{-0.4(M_{tot} - M_{nuc})} . \end{math}
Following \citet{nov06}, we obtain black hole masses for each AGN in our sample using
their relation:
$\log(M(BH)) = 8.19 + 0.524\times(-M_{K (stellar)} - 23)$ or that $\log M(BH) \approx 1.3 \log L(K_{stellar})$.  Our estimated black hole masses can be found in \citet{win08}, along with the 0.3--10\,keV X-ray properties.

We note that aperture photometry in the J and  K band \citep{war8x, gla85} allows one to directly compare the AGN J,K band fluxes to the stellar flux for a sub-sample of our objects. We find that, in general, the total (stellar +nuclear) J and K band flux is dominated by star light with the AGN contribution, even for the most luminous BAT selected AGN, being less than $\approx 30$\%
of the total luminosity. This dominance of the near IR light by stars is beautifully confirmed by the spectroscopic IR data for PG quasars \citep{das07}.  Even in these  very luminous quasars (all of which have non-stellar IR luminosities more than 3 times larger than the stellar luminosities of the most luminous object in our sample and 80 times larger than most of the objects), about 10\% of the total H band light is due to stars. The work of \citet{ben06}, which contains some of the same objects as in our sample, shows that even  in the optical band at 5100\AA \ (where the relative dominance of AGN light  is larger than in the near IR) none of  their objects are dominated by AGN light.   We thus believe that the noise introduced by the measurement error of the  AGN light on the estimate of black hole masses for these objects is relatively small \citep{pen06}.  We have compared our photometry with that in the Peng {\it et al.} paper for both the extended emission and the nuclear emission. Peng {\it et al.} perform a detailed deconvolution of the 2MASS images and thus, presumably have done a more careful job than we have. However we find only $\approx 1/3$ mag scatter in the K band nuclear magnitudes between the Peng results and ours, indicating that our technique is sufficient to the task at hand and our relatively coarse photometry does not effect any of the results in this paper.

\section{Bolometric Correction}
We use a constant ratio of 15 to transfer the Swift BAT (14 -- 195\,keV) luminosity to bolometric luminosity, based on the broad band SEDS \citep{bar05}  and the measurement of an E$^{-2}$ photon index for the Swift BAT sources over the entire 14 -- 195\,keV band. While there seems to be a relation between the softer band x-ray luminosity and bolometric luminosity \citep{ste06}, the effect is rather shallow and we choose to neglect it in this paper since it has not yet been measured in our harder x-ray band. If the effect is similar in the BAT band to that in the 0.5-2\,keV band, it will tend to increase our estimate of the bolometric luminosity of the higher luminosity objects. We are accumulating broad band SEDS for all of these objects, but for many of the objects the nucleus is very weak in the UV and optical bands and thus measuring its luminosity is difficult. To obtain good quality estimates of the nuclear optical and UV fluxes will probably require HST  \citep{mal98} or adaptive optics data.

\begin{figure*}
\plotone{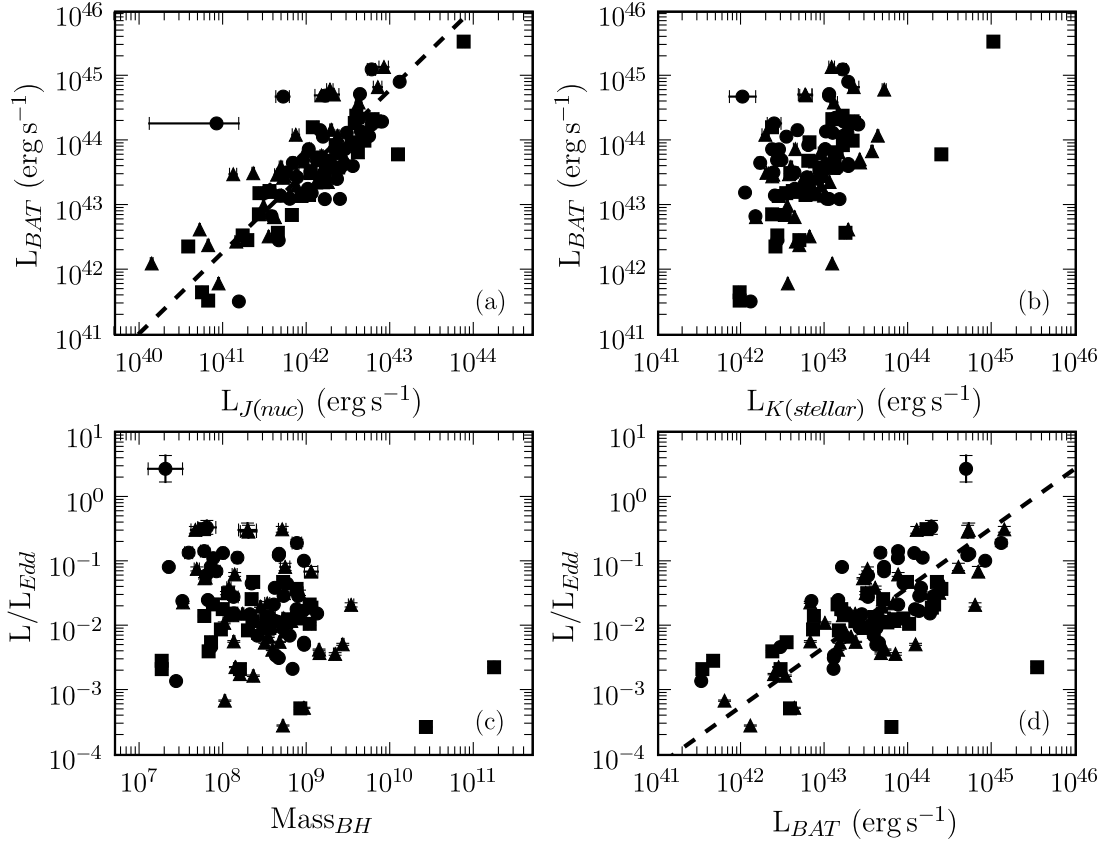}
\caption{Plots of (a) the J filter nuclear luminosity versus the BAT x-ray luminosity (14 -- 195\,keV), (b) the K filter stellar luminosity versus the BAT x-ray luminosity, (c) the
ratio of the bolometric luminosity (estimated from the BAT luminosity) to Eddington luminosity (estimated from M(BH) \citep{nov06} using the K filter stellar (total - nuclear) magnitude) (L/L$_{Edd}$) versus the mass derived from the stellar K filter magnitude (in $M_{\sun}$), and (d) the ratio L/L$_{Edd}$ versus the BAT luminosity.  The circles represent
Sy1 - Sy1.2 sources, the squares represent Sy1.5-Sy1.9, and the triangles represent Sy2s.  For most of the plotted points, the error bars fit entirely within the size of the symbol.  The lines in
panels (a) and (d) represent linear fits to the data using the OLS bisector method (results are discussed in the text).
\label{fig1}}
\end{figure*}

\section{Results}
\subsection{Eddington Ratio}
Using the apparent magnitudes defined in Section 3, we derived associated NIR fluxes
with the flux for zero-magnitude zero point conversion values of $3.129 \times 10^{-10}$\,erg\,s$^{-1}$\,cm$^{-2}$
for the J band and $4.283 \times 10^{-11}$\,erg\,s$^{-1}$\,cm$^{-2}$ for the K$_S$ band \citep{coh03}.
Luminosities, for example the nuclear luminosity L$_{J (nuc)}$ and the galactic/stellar
luminosity L$_{K (stellar)}$, were computed, as for the absolute magnitudes.  Additionally, we computed
the bolometric luminosity (bolometric correction $\times L_{BAT}$) and the Eddington luminosity
(using the estimated black hole mass from $M_{K (stellar)}$ and the \citet{nov06} equation) to
determine the Eddington ratio for each source.

We show in Figure~\ref{fig1} (a) the correlation of BAT x-ray luminosity to L$_{J (nuc)}$, (b) the relation of BAT x-ray luminosity to L$_{K (stellar)}$,
(c) the Eddington ratio versus black hole mass (M(BH)/$M_{\sun}$), and (d) the correlation of the Eddington
ratio to the BAT x-ray luminosity.
 Of the final 3 panels (b,c,d in Figure~\ref{fig1}), only the Eddington ratio to the x-ray luminosity shows a significant correlation, of the form  $\log $L$_{bol}$ /L$_{Edd} = ( 0.924 \pm 0.069)\log$L$_{BAT} + (-42.055 \pm 2.999)$ (solid line in 1d) with a correlation coefficient of R$2$ = 0.45 for 98 points which has a probability of  $3\times10^{-7}$ by chance.
Notice that both absorbed and unabsorbed AGN (primarily type II and I respectively) fall on the same scatter diagram indicating the absence of Compton thick objects, which might be expected to have a high IR to x-ray ratio.
Our results are similar to previous work \citep{wu02} where no correlation of Eddington ratio to black hole mass is found. However, our median Eddington ratio of 0.069 is much less than that found by \citet{kol06} who find values of $\approx 0.2$. We attribute this to the much higher luminosity of the  \citet{kol06} sources, since we find for the most luminous objects in our sample a similar  Eddington ratio of 0.2. The primary relationship is clearly with the BAT hard x-ray luminosity, since our assumption of a constant conversion from BAT to bolometric luminosity can only add noise, as does our method of estimating the black hole mass.

  The prime systematic errors are our estimate of the black hole mass and the bolometric correction. These should, on average,  not  reduce the significance of the relationship between the Eddington ratio and the hard x-ray luminosity. The sense of the errors is to have overestimated the black hole mass due to the improper correction for AGN light (e.g. the black hole mass is maximized in using the Nowak correlations if more of the K or J band light is assumed to be due to stars). This error will be a bigger problem for the higher luminosity objects because more of the K band light in the 2MASS aperture is due to the AGN. Since the hard x-ray luminosity does not have a similar possible error, we may have systematically underestimated the black hole mass and thus overestimated the Eddington ratio in the high luminosity objects. Similarly, if in the BAT band the x-ray light is a smaller fraction of the bolometric luminosity for luminous objects, as it seems to be in the 0.5-2\,keV band for type I objects \citep{ste06}, then the true bolometric luminosity is larger at the higher L(BAT). Both of these errors will tend to reinforce the sense of the correlation.  Alternatively, if we have overestimated the AGN J band light it will reduce the total IR stellar light and thus our mass estimates and effect (see next section) the relationship of IR nuclear to x-ray light. Since this relation  shows  a scatter of only $\approx 0.5$\,dex, this effect does not appear important.
Of course there is some concern (as pointed out by the referee) that this correlation is an artifact of relating  the x-ray luminosity to the x-ray luminosity divided by the black hole mass, where the black hole mass has a small range and the x-ray has a large range. We do not believe that this is the origin of our correlation since the correlation of L(BAT)  with stellar mass has the same narrow range in mass and large range in L(x) and there is no correlation. Almost all correlations with black hole mass have the same bias, since there is only an approximately factor of 300 range in black hole masses estimated from bulge luminosities, one should exercise caution.

\subsection{Correlation of J and K band nuclei luminosities with hard x-ray}
As is obvious from Figure~\ref{fig1}, the luminosity of the nucleus in the BAT and J band is strongly correlated, with a best fit (using the ordinary least square bisector method as presented in \citet{iso86}) of $\log(L_J(nuc)) = \log(L_{BAT})\times(1.257  \pm 0.077) +   (-6.57  \pm  3.09)$ and R$^2 = 0.63$.  The same correlation is seen with the K band nuclear luminosity (not shown).
Using the same OLS bisector method, \citet{ste06} finds a correlation between optical (2500\AA) and soft x-ray (2\,keV) luminosity with a slope of  $0.721 \pm 0.011$ in a sample of unobscured, optically-selected AGN.  The non-linear relationship between optical and soft x-ray emission is partially true due to the fact that low x-ray luminosity sources are more likely to be absorbed \citep{ste03}.
Since the very hard x-ray/BAT and IR luminosities are relatively insensitive to obscuration, the almost linear relationship between these luminosities is not sensitive to obscuration effects and is thus more robust than the optical or UV to soft x-ray correlations. 


\section{Discussion}
The median value of the Eddington ratio that we find is L/L$_{Edd} \approx 0.02$ with a full range of $\approx 1000$.  This value seems to be at odds with the work of recent AGN samples, such as the \citet{kol06} sample selected based on optical/IR/x-ray properties, which finds $L_{Bol}/L_{Edd}$ = 0.25 over a rather different luminosity range $10^{45}$ -- $10^{47}$\,erg\,s$^{-1}$ in L(Bol) than the BAT sample. We believe that this is because the Kollmeier selection is really an optical one, since only the broad line sources have mass estimates.  Such a sample is heavily biased towards ``unobscured'' sources which have higher luminosities \citep{ste03}.  Also, we now know that they also have higher $L/L_{Edd}$ \citep{win08}.  Additionally, the Kollmeier selection is also biased towards much higher redshifts than the BAT survey, which could also explain the difference if, for instance, their is an evolution in Eddington ratios.  The wide range of inferred Eddington ratios from x-ray data has been noticed before \citep{mer05} and is also seen in  other low z samples \citep{hop05}. However, our sample differs significantly from the \citet{ho05} sample in not having objects with very low Eddington ratios. We are not sure why we would have a bias towards higher Eddington ratios than \citet{ho05}. However, since we have not convolved our selection function with a theoretical Eddington ratio model, one cannot be sure that our differences with Ho and Kollmeir are not due to selection biases, the luminosity range covered, and/or evolutionary effects.

Softer x-ray surveys in the 2--10\,keV band (e.g. \citet{ste06}) have found that  the bolometric correction is a function of luminosity, such that the x-ray band carries less of the luminosity at higher luminosities. If this were also true for our sample it would increase the amplitude of the correlation between hard x-ray and bolometric luminosities and change the slope \citep{hop05}.

The correlation between Eddington ratio and x-ray luminosity has, to our knowledge, not been reported from  other samples (e.g. \citet{mer05}) but it is clearly present in them. If one graphs the \citet{mer05} data one finds exactly the same correlation as in our plot. There are theoretical reasons \citep{hop05} to believe that the most luminous objects have the highest Eddington ratios and the lower luminosity objects have lower L$_{Edd}$. The Hopkins paper gives the Eddington ratio as a conditional probability distribution in which the width is a function of Eddington ratio -- this is hard to interpret in our present framework and we have not tested the predictions of this model.  

We note that our distribution of masses is weighted towards higher masses than predicted by \citet{hop05} at low z, but we cannot conclude that this is a strong result  since we have not  convolved  the  \citet{hop05} prediction with our sensitivity limit.

It is well known that there is a break in the luminosity function of AGN, the origin of which is not clear.
Below the break in the luminosity function, the median  Eddington ratio from our sample  is -2.1 and the variance is 0.35,  while at higher  L the median is -1.64 and the variance is 0.199. This is in agreement with the predictions of \citet{hop05}, that the variance in the Eddington ratio should decrease at higher L, and provides a possible explanation of the break in the luminosity function as a change in the mean Eddington ratio of the population.

Our finding of a strong almost linear correlation of the near IR luminosities with the BAT luminosity (see \citet{hor07} for a similar result) argues that similar processes produce the near IR continuum and the hard x-rays. Since the mechanism producing the x-ray emission is unrelated to dust, these results do not favor dust reprocessing models for the origin of the near IR continuum, unless the main driver of the dust reprocessing is simply and linearly connected to the x-ray continuum intensity.

\acknowledgments
DHM acknowledges support from the National Aeronautics and Space
Administration (NASA) under LTSA Grant NAG5-13102 issued through the
Office of Space Science.
\\
This publication makes use of data products from the Two Micron All Sky Survey.
\\
We acknowledge the work that the Swift BAT team has done which have made this work possible.
{\it Facilities:} \facility{Swift (BAT)}


\begin{thebibliography}{}
\bibitem[Alonso-Herrero {\it et al.}(1997)]{alo9x} Alonso-Herrero, A. {\it et al.} 1997, \mnras, 288, 977
\bibitem[Alonso-Herrero {\it et al.}(2001)]{alo0x}Alonso-Herrero, A. {\it et al.} 2001, AJ, 121, 1369
\bibitem[Barger {\it et al.}(2005)]{bar05} Barger, A.J. {\it et al.} 2005, \aj, 129, 578
\bibitem[Barmby {\it et al.}(2006)]{bar06} Barmby, P. {\it et al.} 2006, \apj, 642, 126
\bibitem[Bentz {\it et al.}(2006)]{ben06} Bentz, M.C. {\it et al.} 2006, \apj, 644,142
\bibitem[Boroson(2002)]{bor0x} Boroson, T.A. 2002, \apj, 565, 78
\bibitem[Brocksopp {\it et al.}(2006)]{bro06} Brocksopp, C. {\it et al.} 2006, \mnras, 366, 953
\bibitem[Cohen {\it et al.}(2003)]{coh03} Cohen, M. {\it et al.} 2003, AJ,126, 1090
\bibitem[Czerny \& Elvis(1987)]{cze87} Czerny, B. \& Elvis, M. 1987, \apj, 321, 305
\bibitem[Dasyra {\it et al.}(2007)]{das07} Dasyra, K.M. {\it et al.} 2007, \apj, 657, 102
\bibitem[Elvis {\it et al.}(1994)]{elv94} Elvis, M. 1994, \apjs, 95, 1
\bibitem[Franceschini {\it et al.}(2005)]{fra05} Franceschini, A. {\it et al.} 2005, \aj, 129,2074
\bibitem[Glass \& Moorwood(1985)]{gla85} Glass, I. S. \& Moorwood, A. F. M. 1985, \mnras, 214, 429
\bibitem[Glass(2004)]{gla04} Glass, I.S. 2004, \mnras, 350, 1049
\bibitem[Ho, Filippenko, \& Sargent(1995)]{ho05} Ho, L. C., Filippenko, A. V., \& Sargent, W. L. 1995, \apjs, 98, 477
\bibitem[Hopkins, Narayan, \& Hernquist(2005)]{hop05} Hopkins, P.F., Narayan, R., \& Hernquist, L. 2006, \apj,  643, 641
\bibitem[Hopkins {\it et al.}(2006)]{hop06} Hopkins, P.F. {\it et al.} 2006, \apj, 652, 864
\bibitem[Horst, {\it et al.}(2007)]{hor07} Horst, H., {\it et al.} 2007, astro-ph/0711.3734
\bibitem[Isobe, Feigelson, \& Nelson(1986)]{iso86} Isobe, T., Feigelson, E. D., \& Nelson, P. I. 1986, \apj, 306, 490
\bibitem[Kollmeier {\it et al.}(2006)]{kol06} Kollmeier, J. 2006, \apj, 648, 128
\bibitem[K\"{o}rding, Falcke, \& Corbel(2006)]{kor06} K\"{o}rding, E., Falcke, H., \& Corbel, S. 2006, A\&A, 456, 439
\bibitem[Krabbe, B\"{o}ker, \& Maiolino(2001)]{kra01} Krabbe, A., B\"{o}ker, T., \& Maiolino, R. 2001, \apj, 557, 626
\bibitem[La Franca {\it et al.}(2005)]{laf05} La Franca, F. {\it et al.} 2005, \apj, 635, 864
\bibitem[Lutz {\it et al.}(2004)]{lut04} Lutz, D. {\it et al.} 2004, A\&A, 418, 465
\bibitem[Malkan {\it et al.}(1998)]{mal98} Malkan, M.A. {\it et al.} 1998, \apjs, 117, 25
\bibitem[Merloni, Heinz, \& di Matteo(2005)]{mer05} Merloni, A., Heinz, S., \& di Matteo, T. 2003, \mnras, 345,1057
\bibitem[Markwardt {\it et al.}(2005)]{mar05} Markwardt, C.B. {\it et al.} 2005, \apj, 633, 77
\bibitem[Matt {\it et al.}(2006)]{mat06} Matt, G. 2006, MmSAI, 77, 606
\bibitem[Nenkova {\it et al.}(2008)]{nen08} Nenkova, M., {\it et al.} 2008, ApJ (accepted) astro-ph/0806.0511
\bibitem[Netzer \& Trakhtenbrot(2007)]{net07} Netzer, H. \& Trakhtenbrot, B. 2007,\apj, 654, 754
\bibitem[Novak, Faber, \& Dekel(2006)]{nov06} Novak, G.S., Faber, S. M., \& Dekel, A. 2006, \apj, 637, 96
\bibitem[Peng {\it et al.}(2006)]{pen06} Peng, Z.,  Gu, Q., Melnick, J., \& Zhao, Y. 2006, A\&A, 453, 863
\bibitem[Pier \& Krolik(1993)]{pie93} Pier, E.A. \& Krolik, J.H. 1993, \apj, 418, 673
\bibitem[Richards {\it et al.}(2006)]{ric06} Richards, G.T. {\it et al.} 2006, \apjs, 166, 470
\bibitem[Steffen {\it et al.}(2003)]{ste03} Steffen, A.T.  {\it et al.} 2003, \apj, 596, 23
\bibitem[Steffen {\it et al.}(2006)]{ste06} Steffen, A.T. {\it et al.} 2006, AJ, 131, 2826
\bibitem[Suganuma {et al.}(2006)]{sug05} Suganuma, M. {\it et al.} 2006, \apj, 639, 46
\bibitem[Sun \& Malkan(1989)]{sun89} Sun, W. \& Malkan, M.A. 1989, \apj, 346, 68
\bibitem[Tozzi {\it et al.}(2006)]{toz06} Tozzi, P. {\it et al.} 2006, A\&A, 451, 457
\bibitem[Tueller {\it et al.}(2008)]{tue07} Tueller, J. {\it et al.} 2008, \apj, 681, 113
\bibitem[Ward {\it et al.}(1988)]{war8x} Ward, M.J. {\it et al.} 1988, \apj, 324, 767
\bibitem[Winter {\it et al.}(2008)]{win08} Winter, L.M. {\it et al.} 2008, \apj \,(submitted)
\bibitem[Woo \& Urry(2002)]{wu02} Woo, J.-H. \& Urry, C.M. 2002, \apj, 581, 5
\bibitem[Zdziarski {\it et al.}(1990)]{zdz9x} Zdziarski, A.A. {\it et al.} 1990, \apj, 363, 1
\end{thebibliography}
\end{document}